\documentclass[twocolumn,aps,showpacs]{revtex4}
\usepackage[utf8]{inputenc}
\usepackage{color}
\usepackage{amsmath}
\usepackage{graphicx}

\makeatletter
\@ifundefined{textcolor}{}
{%
 \definecolor{BLACK}{gray}{0}
 \definecolor{WHITE}{gray}{1}
 \definecolor{RED}{rgb}{1,0,0}
 \definecolor{GREEN}{rgb}{0,1,0}
 \definecolor{BLUE}{rgb}{0,0,1}
 \definecolor{CYAN}{cmyk}{1,0,0,0}
 \definecolor{MAGENTA}{cmyk}{0,1,0,0}
 \definecolor{YELLOW}{cmyk}{0,0,1,0}
}

\usepackage{dcolumn}
\usepackage{bm}
\usepackage{wrapfig}
\usepackage{subfigure}
\usepackage{ucs}
\usepackage{lineno}
\usepackage{epsfig}
\usepackage{psfrag}\usepackage{epstopdf}
\DeclareGraphicsExtensions{.pdf,.eps,.png,.jpg,.mps}

\makeatother

\begin{document}
\title{Thermal entanglement and correlated coherence in two coupled double
quantum dots systems}
\author{Cleverson Filgueiras$^{1}$, Onofre Rojas$^{1}$,
Moises Rojas$^{1}$}
\affiliation{$^{1}$Departamento de Física, Universidade Federal de Lavras, 37200-900,
Lavras-MG, Brazil}
\begin{abstract}
In this work, we investigate the thermal quantum correlations in two coupled double semiconductor charge qubits. This is carried out by deriving analytical expressions for both the thermal concurrence and the correlated coherence. We study, in detail, the effects of the tunneling parameters, the Coulomb interaction and the temperature on the thermal entanglement and on the correlated
coherence. It is found that the Coulomb potential plays an important role in the thermal entanglement and in the correlated coherence of the system. The results also indicate that the Coulomb potential can be used for significant enhancement of the thermal entanglement and quantum coherence. One interesting aspect is that the correlated
coherence capture all the thermal entanglement at low temperatures, i.e,  the local coherences are totally transferred to
the thermal entanglement. Finally, we focus on the role played by thermal entanglement and the correlated coherence responsible for quantum correlations. We show that in all cases, the correlated coherence is more robust than the thermal entanglement so that quantum algorithms based only on correlated coherence may be more robust than those based on entanglement. Our results also show that the entanglement can be tuned by varying the Coulomb interaction between electrons.
\end{abstract}
\maketitle
\section{introduction}
The entanglement carries not only interesting properties of quantum mechanics
but also it is a very important phenomenon due to the powerful applications in the context of quantum
information process and quantum computing\cite{Ben1,Ben2,lamico}. In the last decades, solid-state realizations of it have received considerable attention due to the fact that semiconductor nanostructures
such as quantum dots \cite{peta,press} and double quantum dots (also known as quantum dot molecules) \cite{shin,aus}
are promising candidates for the physical implementation of quantum information
processing\cite{so}. There are proposals for quantum dots using either
charge \cite{gor} or spin \cite{benito,loss,an} as qubits, or even both at the same time
\cite{shi,yang}. These quantum systems are of great interest due to their
easy integration with the existing electronics and the advantage of scalability
\cite{ita,urda}. The coherent control of tunneling in a asymmetric
double quantum dot was reported in \cite{villas}. Moreover, the
quantum dynamics and the entanglement of two electrons inside the coupled
double quantum dots were addressed in \cite{sanz,sza}, while that the aspects related
to the quantum correlations and to the decoherence were investigated in \cite{fan,qin,borge,sou}. In addition, quantum teleportation based on the double quantum dots \cite{choo}, the quantum noise due to phonons induce steady-state
in a double quantum dot charge qubit\cite{gia},  multielectron
quantum dots \cite{rao} and  spin-orbit-coupled quantum memory of a double quantum dot \cite{chot} were also reported.

On the other hand, quantum coherence is one of the central concepts
in quantum mechanical systems. By arising from the quantum
superposition, it is a fundamental feature of many different quantum mechanical
systems such as electrons, photons, atoms, hybrid systems and so on.
It has been widely used as a resource in the quantum information processing
\cite{stre1}, in the quantum metrology \cite{fro,gio}, in the thermodynamics \cite{brandao,lan}, etc. Several measurements of it were proposed, so its
properties have been investigated in detail over the years(see \cite{baum,Hu}, for instance). More recently, a new measure called correlated coherence \cite{tan,tri}
was introduced in order to investigated the relation between the quantum coherence
and the quantum correlations. The quantum correlated coherence is a measure of coherence with the local parts removed, is to say, all coherence in the system is stored entirely within the quantum correlations.

In this paper, our main goal is to investigate the thermal entanglement
and the quantum correlated coherence in two couples quantum dot molecules,
where are regarded an isolated double quantum dot as a charge qubit. We take into account that the system is isolated from their respective
electronic reservoirs and that it remains in the strong Coulomb blockade regime,
such that one excess electron is permitted in each double
quantum dot, at most. In our model, we consider a system of two charge qubits,
in which an excess electron occupies either the left dot $|L\rangle$ or the
right one, $|R\rangle$. We obtained analytical solutions, which allowed us to explore in detail the performance of the thermal entanglement.
We also derived the quantum correlated coherence and we investigated the
role played by it in our model. In addition, it is compared
the thermal entanglement with a quantum correlated coherence. Finally, the framework provided by the correlated coherence allows us to retrieve the same concepts of quantum discord as well as quantum entanglement, providing a unified view of these correlations, where the quantum discord is a measure of the quantum correlations going beyond entanglement \cite{ollivier,wer}. Note that, for a multipartite system, if the coherence of the global state is  resource which cannot be increased, the cost of creating discord can be expressed in terms of coherence \cite{yue,ma}.

This paper is organized as follows. In Section II we describe the
physical model and the method treat it. In Section III, we
briefly review the definition of the concurrence $\mathcal{C}$ and the correlated
coherence, $\mathcal{C}_{cc}$. The analytical expressions 
for them are found. In the Section
IV, we discuss the most interesting results of the behavior the thermal
entanglement and the correlated coherence. Finally, in Section V, we summarize our main conclusions. 
\section{The model}
\begin{figure}
\includegraphics[scale=0.55]{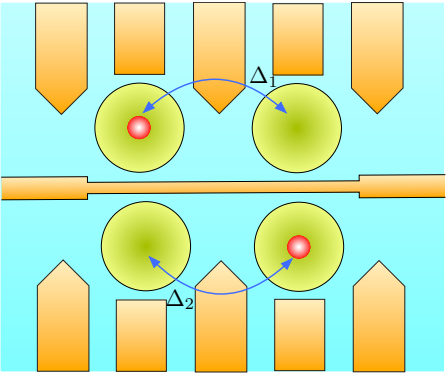}\caption{\label{fig:model}(Color online) A schematic representation of the
physical model with the two coupled double quantum dots. The electrons are represented by the smaller spheres inside the quantum dots.}
\end{figure}

The model consists in the two sets of double quantum dots (DQDs), where
each dot is filled with a single electron, in which an excess electron
is located either in the left dot($\left|L\right>$) or in the right one($\left|R\right>$), as shown in Fig. \ref{fig:model}.
The Hamiltonian of the two couple double quantum dots \cite{fan}
is given by
\begin{equation}
\begin{array}{ccc}
H & = & \Delta_{1}\sigma_{1}^{x}+\Delta_{2}\sigma_{2}^{x}+V\left(\sigma_{1}^{z}\otimes\sigma_{2}^{z}\right)\end{array}\label{eq:1}
\end{equation}
where $\sigma_{1(2)}^{\alpha}(\alpha=x,z)$ are the Pauli operators,
$\Delta_{1(2)}$ is the strength of the tunneling coupling between
the two quantum dots, while the $V$ represents the Coulomb interactions
between the electrons. 
We consider the convention $\left|0\right>\equiv\left|L\right>$ and
$\left|1\right>\equiv\left|R\right>$ to indicate the electron occupying
either the left dot, $\left(L\right)$, or the right one, $\left(R\right)$.
The four eigenvectors of Eq. (\ref{eq:1}) in the standard basis $\left\{ \left|LL\right>,\left|LR\right>,\left|RL\right>,\left|RR\right>\right\} $
are

\begin{eqnarray}
\left|\varphi_{1}\right> & = & \alpha_{-}\left[A_{-}\left(-\left|LL\right>+\left|RR\right>\right)+n_{-}\left(\left|LR\right>-\left|RL\right>\right)\right],\nonumber \\
\left|\varphi_{2}\right> & = & \alpha_{-}\left[n_{-}\left(-\left|LL\right>+\left|RR\right>\right)+A_{-}\left(-\left|LR\right>+\left|RL\right>\right)\right],\nonumber \\
\left|\varphi_{3}\right> & = & \alpha_{+}\left[A_{+}\left(\left|LL\right>+\left|RR\right>\right)+n_{+}\left(\left|LR\right>+\left|RL\right>\right)\right],\nonumber \\
\left|\varphi_{4}\right> & = & \alpha_{+}\left[n_{+}\left(\left|LL\right>+\left|RR\right>\right)-A_{+}\left(\left|LR\right>+\left|RL\right>\right)\right].
\end{eqnarray}
where $\alpha_{\pm}=\frac{1}{\sqrt{2}\sqrt{(n_{\pm})^{2}+A_{\pm}^{2}}}$
, $A_{\pm}=V+\sqrt{(n_{\pm})^{2}+V^{2}}$ , $n_{\pm}=\Delta_{1}\pm\Delta_{2}$
and the corresponding eigenvalues are

\begin{eqnarray}
\varepsilon_{1,2} & = & \pm\sqrt{(n_{-})^{2}+V^{2}},\\
\varepsilon_{3,4} & = & \pm\sqrt{(n_{+})^{2}+V^{2}}.
\end{eqnarray}

The system state in the thermal equilibrium is described by $\rho(T)=\frac{\exp(-\beta H)}{Z}$,
where $\beta=1/k_{B}T$, with $k_{B}$ being the Boltzmann's constant,
$T$ is the absolute temperature and the partition function of the
system is defined by $Z=Tr\left[\exp(-\beta H)\right]$.
\subsection{The density operator}
The state of this bipartite system at the thermal equilibrium can be expressed
by the density operator $\rho$ as
\begin{equation}
\rho_{AB}(T)=\left[\begin{array}{cccc}
\rho_{11} & \rho_{12} & \rho_{13} & \rho_{14}\\
\rho_{12} & \rho_{22} & \rho_{23} & \rho_{13}\\
\rho_{13} & \rho_{23} & \rho_{22} & \rho_{12}\\
\rho_{14} & \rho_{13} & \rho_{12} & \rho_{11}
\end{array}\right].\label{eq:5}
\end{equation}
The elements of this density matrix, after an cumbersome algebraic
manipulation, are given by 

\[
\begin{array}{ccc}
\rho_{11} & = & \frac{\alpha_{-}^{2}\left(A_{-}^{2}e^{-\beta\varepsilon_{1}}+n_{-}^{2}e^{-\beta\varepsilon_{2}}\right)+\alpha_{+}^{2}\left(A_{+}^{2}e^{-\beta\varepsilon_{3}}+n_{+}^{2}e^{-\beta\varepsilon_{4}}\right)}{Z},\\
\rho_{12} & = & \frac{A_{-}n_{-}\alpha_{-}^{2}\left(-e^{-\beta\varepsilon_{1}}+e^{-\beta\varepsilon_{2}}\right)+A_{+}n_{+}\alpha_{+}^{2}\left(e^{-\beta\varepsilon_{3}}-e^{-\beta\varepsilon_{4}}\right)}{Z},\\
\rho_{13} & = & \frac{A_{-}n_{-}\alpha_{-}^{2}\left(e^{-\beta\varepsilon_{1}}-e^{-\beta\varepsilon_{2}}\right)+A_{+}n_{+}\alpha_{+}^{2}\left(e^{-\beta\varepsilon_{3}}-e^{-\beta\varepsilon_{4}}\right)}{Z},\\
\rho_{14} & = & \frac{-\alpha_{-}^{2}\left(A_{-}^{2}e^{-\beta\varepsilon_{1}}+n_{-}^{2}e^{-\beta\varepsilon_{2}}\right)+\alpha_{+}^{2}\left(A_{+}^{2}e^{-\beta\varepsilon_{3}}+n_{+}^{2}e^{-\beta\varepsilon_{4}}\right)}{Z},\\
\rho_{22} & = & \frac{\alpha_{-}^{2}\left(n_{-}^{2}e^{-\beta\varepsilon_{1}}+A_{-}^{2}e^{-\beta\varepsilon_{2}}\right)+\alpha_{+}^{2}\left(n_{+}^{2}e^{-\beta\varepsilon_{3}}+A_{+}^{2}e^{-\beta\varepsilon_{4}}\right)}{Z},\\
\rho_{23} & = & \frac{-\alpha_{-}^{2}\left(n_{-}^{2}e^{-\beta\varepsilon_{1}}+A_{-}^{2}e^{-\beta\varepsilon_{2}}\right)+\alpha_{+}^{2}\left(n_{+}^{2}e^{-\beta\varepsilon_{3}}+A_{+}^{2}e^{-\beta\varepsilon_{4}}\right)}{Z},
\end{array}
\]
where $Z=\underset{i}{\sum}e^{-\beta\varepsilon_{i}}$.

Since $\rho_{AB}(T)$ represents a thermal state, the entanglement is
then called {\it thermal entanglement}. 
\section{Quantum Correlations}
\subsection{Thermal entanglement }

In order to quantify the thermal entanglement, we employ the quantity called concurrence $\mathcal{C}$
\cite{wootters}. Which is defined as
\begin{eqnarray}
\mathcal{C}={\rm {max}\left\{ 0,\mid\sqrt{\lambda_{1}}-\sqrt{\lambda_{3}}\mid-\sqrt{\lambda_{2}}-\sqrt{\lambda_{4}}\right\} ,}
\end{eqnarray}
where $\lambda_{i}\:(i=1,2,3,4)$ are the eigenvalues in the decreasing
order of the matrix 
\begin{eqnarray}
R=\rho\left(\sigma^{y}\otimes\sigma^{y}\right)\rho^{\ast}\left(\sigma^{y}\otimes\sigma^{y}\right),
\end{eqnarray}
with $\sigma^{y}$ being the Pauli matrix. We will explore the thermal
entanglement of the considered two coupled double quantum dots systems.
To achieve this goal, we must obtain the eigenvalues of $R$. They are given by

\begin{eqnarray}
\lambda_{1} & = & \frac{\Theta_{-}}{2}+\frac{1}{2}\sqrt{\Xi_{-}^{2}-\Sigma_{-}^{2}},\nonumber \\
\lambda_{2} & = & \frac{\Theta_{-}}{2}-\frac{1}{2}\sqrt{\Xi_{-}^{2}-\Sigma_{-}^{2}},\nonumber \\
\lambda_{3} & = & \frac{\Theta_{+}}{2}+\frac{1}{2}\sqrt{\Xi_{+}^{2}-\Sigma_{+}^{2}},\nonumber \\
\lambda_{4} & = & \frac{\Theta_{+}}{2}-\frac{1}{2}\sqrt{\Xi_{+}^{2}-\Sigma_{+}^{2}},
\end{eqnarray}
where
\[
\begin{array}{ccl}
\Xi_{\pm} & = & \left(\rho_{11}\pm\rho_{14}\right)^{2}-\left(\rho_{22}\pm\rho_{23}\right)^{2},\\
\varSigma_{\pm} & = & 2\left(\rho_{13}\pm\rho_{14}\right)\left(\mp\rho_{11}-\rho_{14}+\rho_{23}\pm\rho_{22}\right),\\
\Theta_{\pm} & = & \left(\rho_{11}\pm\rho_{14}\right)^{2}-2\left(\rho_{12}\pm\rho_{13}\right)^{2}+\left(\rho_{22}\pm\rho_{23}\right)^{2}.
\end{array}
\]

In this case, the analytical expression for the thermal concurrence
is too large to be explicitly provided in this paper. 
\subsection{Correlated Coherence}
The quantum coherence is an useful resource for the quantum information processing task. When in a bipartite system, it can be contained either locally or in the correlations between the subsystems. Its portion for which all the coherence in the system is stored, entirely within the quantum correlations, is called {\it correlated coherence},  $\mathcal{C}_{cc}$ \cite{tan}. For a bipartite quantum system, it becomes
\begin{equation}
\mathcal{C}_{cc}(\rho_{AB})=\mathcal{C}_{l_{1}}(\rho_{AB})-\mathcal{C}_{l_{1}}(\rho_{A})-\mathcal{C}_{l_{1}}(\rho_{B}),\label{eq:6}
\end{equation}
where $\rho_{A}=Tr_{B}(\rho_{AB})$ and $\rho_{B}=Tr_{A}(\rho_{AB})$.
Here, $A$ and $B$ stand for local subsystems.

In accordance with the set of properties which every proper measure of coherence should satisfy \cite{baum}, a number of coherence measures have been put forward. We focus on the $l_{1}$-norm of coherence $\mathcal{C}_{l_{1}}$. It is defined as

\begin{equation}
\mathcal{C}_{l_{1}}(\rho)=\sum_{i\neq j}|\langle i|\rho|j\rangle|.
\end{equation}

The quantum coherence is a basis dependent concept, but we can choose an incoherent one for the local coherence, which will enable us to diagonalize $\rho_{A}$ and $\rho_{B}$. From Eq.(\ref{eq:5}), the reduced density matrix $\rho_{A}(T)$ will be given by
\begin{equation}
\rho_{A}(T)=\left(\begin{array}{cc}
\rho_{11}+\rho_{22} & 2\rho_{13}\\
2\rho_{13} & \rho_{11}+\rho_{22}
\end{array}\right).\label{eq:rha}
\end{equation}
In a similar way, we obtain
\begin{equation}
\rho_{B}(T)=\left(\begin{array}{cc}
\rho_{11}+\rho_{22} & 2\rho_{12}\\
2\rho_{12} & \rho_{11}+\rho_{22}
\end{array}\right).\label{eq:rhb}
\end{equation}
In order to analyze the correlated coherence, we make an unitary transformation in the reduced density matrix $\rho_{A}(T)$ and $\rho_{B}(T)$, for which
\begin{equation}
U=\left(\begin{array}{cc}
\cos\theta & -e^{i\varphi}\sin\theta\\
e^{-i\varphi}\sin\theta & \cos\theta
\end{array}\right),\label{eq:7}
\end{equation}
$\widetilde{\rho}_{A}(T)=U\,\rho_{A}(T)\,U^{\dagger}$ and $\widetilde{\rho}_{B}(T)=U\,\rho_{B}(T)\,U^{\dagger}$. On the other hand, the unitary transformation of the
bipartite quantum state $\rho_{AB}(T)$ is given by $\widetilde{\rho}_{AB}(T)=\widetilde{U}\,\rho_{AB}(T)\,\widetilde{U}^{\dagger}$, where $\widetilde{U}=U\otimes U$. 

The unitary transformation will show the relationship
between the global coherence and the local coherence for several choices of the parameters $\theta$ and $\varphi$. In particular, by setting $(\theta=\frac{\pi}{4},\varphi=0)$ in the Eq.(\ref{eq:7}), we obtain a matrix that diagonalize $\rho_{A}(T)$ and $\rho_{B}(T)$. This step provide us the basis set, where $A$ and $B$ are locally incoherent. Thus, by inserting Eq.(\ref{eq:7}) into the Eq. (\ref{eq:6}), fixing $\theta=\frac{\pi}{4}$
and $\varphi=0$, we obtain an explicit expression for correlated
coherence, that is, 
\begin{equation}
\mathcal{C}_{cc}(\rho_{AB}(T))=|\rho_{11}+\rho_{14}-\rho_{22}-\rho_{23}|+|\rho_{11}-\rho_{14}-\rho_{22}+\rho_{23}|.
\end{equation}

\section{results and discussions}
\begin{figure}
\includegraphics[scale=0.4]{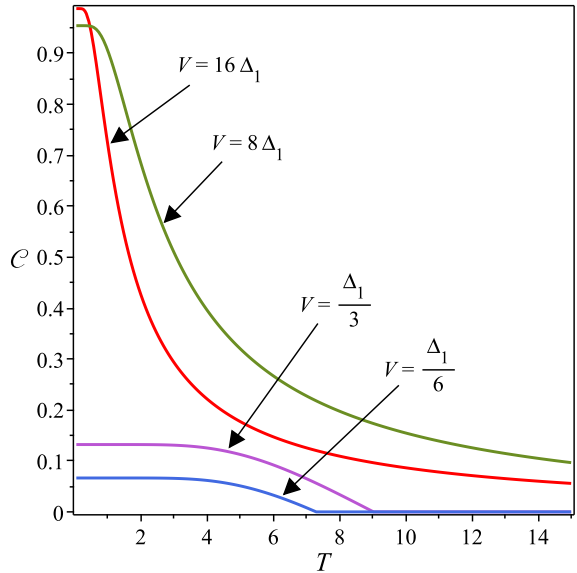}\caption{\label{fig:C-T} The concurrence $\mathcal{C}$ as a function of temperature
$T$ and for various values of the Coulomb potential $V$, with $\Delta_{1}=10$ and $\Delta_{2}=15$.}
\end{figure}

In this section, it is discussed the main results obtained
in the foregoing section. In Fig. \ref{fig:C-T} we show the concurrence $\mathcal{C}$ as a
function of the temperature $T$ and for different values of the potential $V$, with $\Delta_{1}=10$ and $\Delta_{2}=15$. We consider two different regimes: in the first one, for a strong Coulomb potential and for a fixed $V=16\Delta_{1}$
(red curve), we can see that the concurrence for $T=0$ is $\mathcal{C}\approx0.988$. It is observed that the concurrence monotonously tends to
zero as soon as the temperature increases. For $V=8\Delta_{1}$ (green curve), the concurrence is slightly smaller than to the previous case. However, it decreases slower when the temperature is raised. 
In the second regime, we consider a weak Coulomb potential, with $V=\frac{\Delta_{1}}{3}$ (magenta curve)  and $V=\frac{\Delta_{1}}{6}$ (blue curve). In these situations, we have a weak entanglement at zero temperature, remaining almost
constant at low temperatures. As the temperature is raised, it vanishes for $T=9$ and $T=7.2$, respectively. It is interesting to analyze the strong potential regime ($V=16\Delta_{1}$), for which we calculated each element of the density operator in the limit $T\rightarrow0$, getting $\left|\varphi_{4}\right>\approx0.055\left(\left|LL\right>+\left|RR\right>\right)-0.705\left(\left|LR\right>+\left|RL\right>\right)$,
which will generate the concurrence $\mathcal{C}\approx0.988$. On
the other hand, by analyzing $V=\frac{\Delta_{1}}{6}$(the weak potential regime) as $T\rightarrow0$, a single non-zero
state vetor is obtained which contributes to the density operator. This way, we have $\left|\varphi_{4}\right>\approx0.483\left(\left|LL\right>+\left|RR\right>\right)-0.516\left(\left|LR\right>+\left|RL\right>\right)$, thus the concurrence becomes $\mathcal{C}\approx0.066$.

\begin{figure}
\includegraphics[scale=0.4]{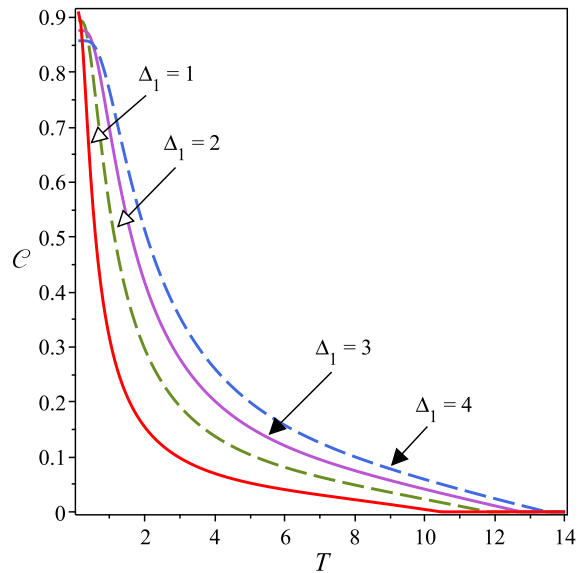}\caption{\label{fig:C-T1} The concurrence $\mathcal{C}$ as a function of
temperature $T$ and for different values of $\Delta_{1}$. Here, $V=20$ and $\Delta_{2}=8$.}
\end{figure}

In Fig. \ref{fig:C-T1}, it is depicted the behavior of the concurrence $\mathcal{C}$ as a function of the temperature $T$ with fixed parameter values ($V=20,\,\Delta_{2}=8$) and for different values of the $\Delta_{1}$. The curves of the concurrence are quite similar to the previous examined case, but
the value of $\mathcal{C}$ in $T=0$ diminishes as the parameter $\Delta_{1}$ increases. In this figure, it can be noted that when
the tunneling parameter is weak (i.e. $\Delta_{1}=1$), the concurrence for $T=0$ is $\mathcal{C}\approx0.912$. And, as $T$ increases, the concurrence $\mathcal{C}$ decreases rapidly until reaching the threshold temperature, above which the thermal entanglement becomes null. As
the tunneling parameter ($\Delta_{1}$) increases, the concurrence
exhibits a similar behavior, but the concurrence is slightly more robust in these cases.

\begin{figure}
\includegraphics[scale=0.4]{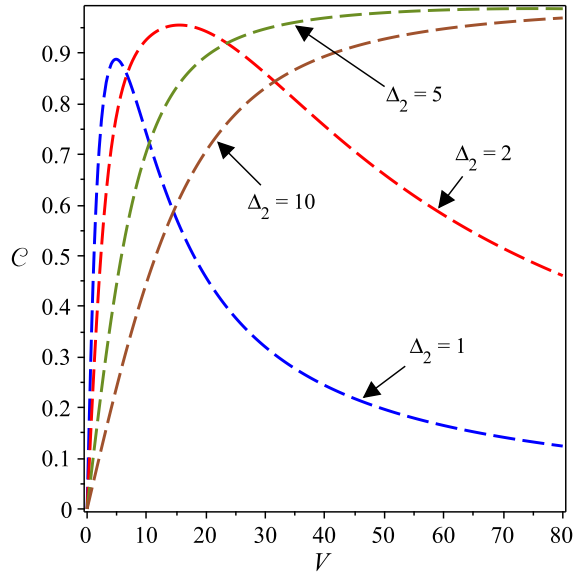}\caption{\label{fig:C-V} The concurrence $\mathcal{C}$ as a function of the
Coulomb potential $V$ for several values of coupling $\Delta_{2}$,
with $\Delta_{2}=\Delta_{1}$ and $T=0.1$. }
\end{figure}

In Fig. \ref{fig:C-V}, we plot the concurrence as a function of
Coulomb potential $V$ for several values of tunneling parameter
$\Delta_{2}$ and fixed $T=0.1$. In this case, we consider the same
tunneling parameter for each double quantum dot. For weak tunneling
parameter $\Delta_{2}=1$, we observe a sudden increase of the concurrence, which attains the value $\mathcal{C}\approx0.9$, which decreases gradually to zero as $V$ increases. Moreover, the figure shows that the thermal entanglement between the charges qubits is more robust as the tunneling parameter is raised (see curves for $\Delta_{2}=2,\,5,\,10$ ). Furthermore, in this figure, it is observed that the concurrence is null at $V=0$ for each parameter $\Delta_{2}$ considered. By analyzing the matrix elements of the density operator, it is observed that the eigenstate is given by $\left|\varphi_{4}\right>=\frac{1}{2}\left(\left|LL\right>+\left|RR\right>\right)-\frac{1}{2}\left(\left|LR\right>+\left|RL\right>\right)$. It is worth to mention that this state is an unentangled one.
This result shows that $V$ can be used for either turning on or off the entanglement. The physical explanation of our model exhibit high entanglement is due to the strong coupling of the Coulomb potential, since when two electrons are relativity close to each other, the Coulomb's interaction energy is greater than it is when the electrons are furthest apart, strongly favoring the entanglement of the left electron of the top double quantum dot and the right electron of the bottom double quantum dot, and vice versa. On the other hand, when the double quantum dots are physically separated, the effects of the Coulomb potential is rather weak, in this case, the probability of finding the electrons is distributed between states $\left\{\left|LL\right>,\left|LR\right>,\left|RL\right>,\left|RR\right>\right\}$, which correspond to the system having weak entanglement. In the limiting case, when the Coulomb potential is null, the electrons will be equally distributed between states  $\left\{\left|LL\right>,\left|LR\right>,\left|RL\right>,\left|RR\right>\right\}$, thus the thermal entanglement is null. In conclusion, we can use the Coulomb potential to tune in the entanglement of the system.
\begin{figure}
\includegraphics[scale=0.36]{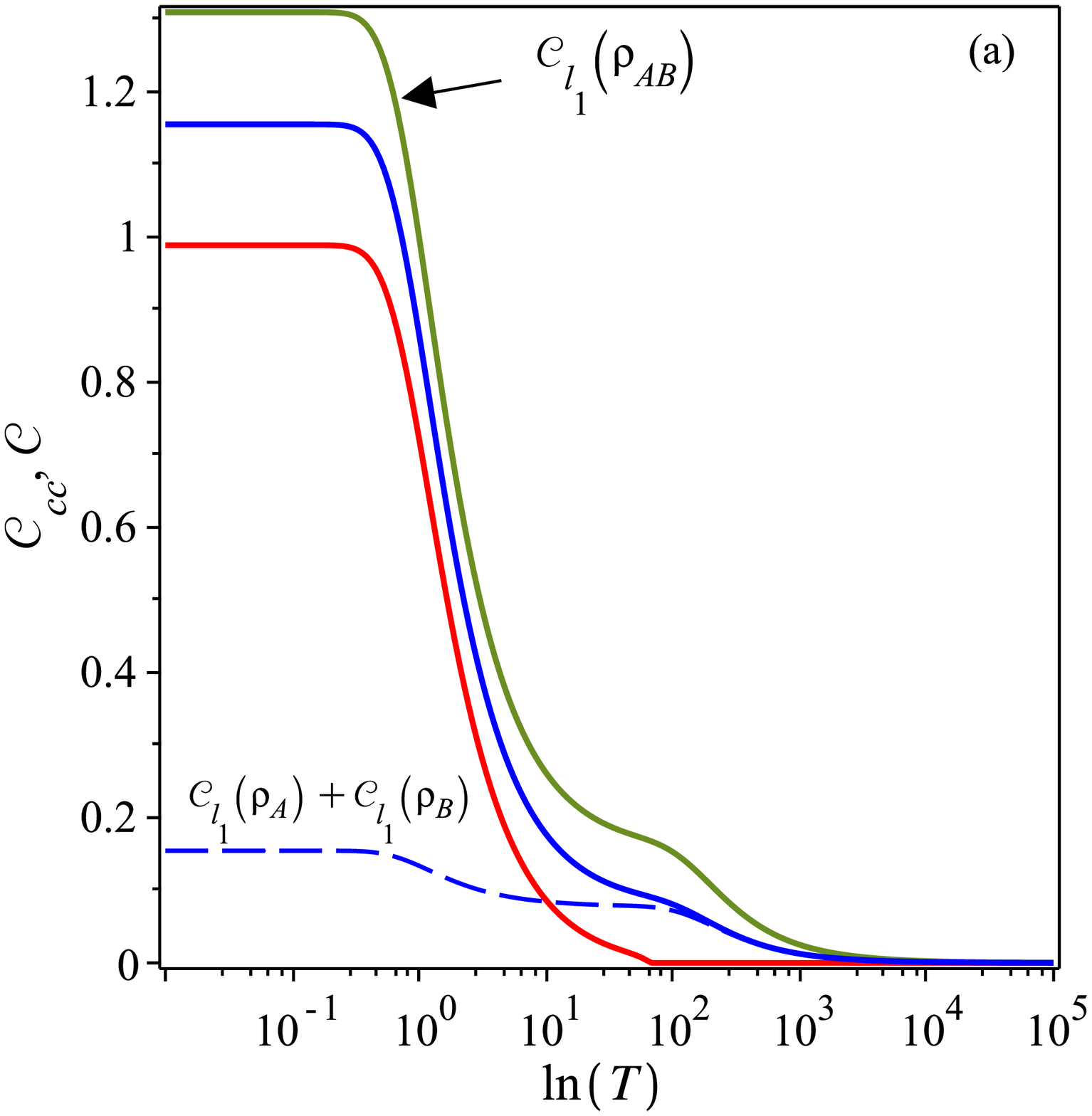}

\includegraphics[scale=0.36]{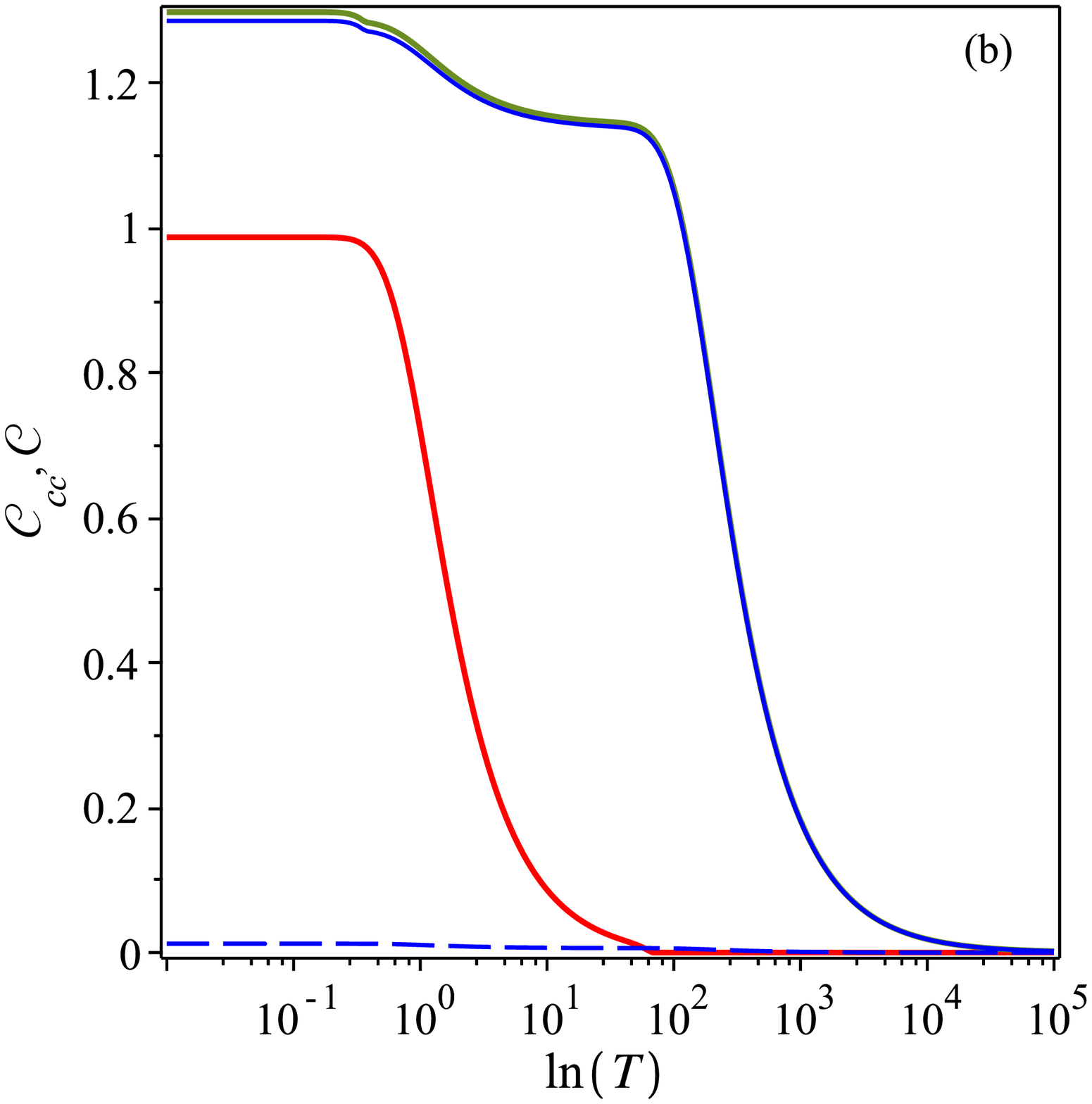}

\includegraphics[scale=0.36]{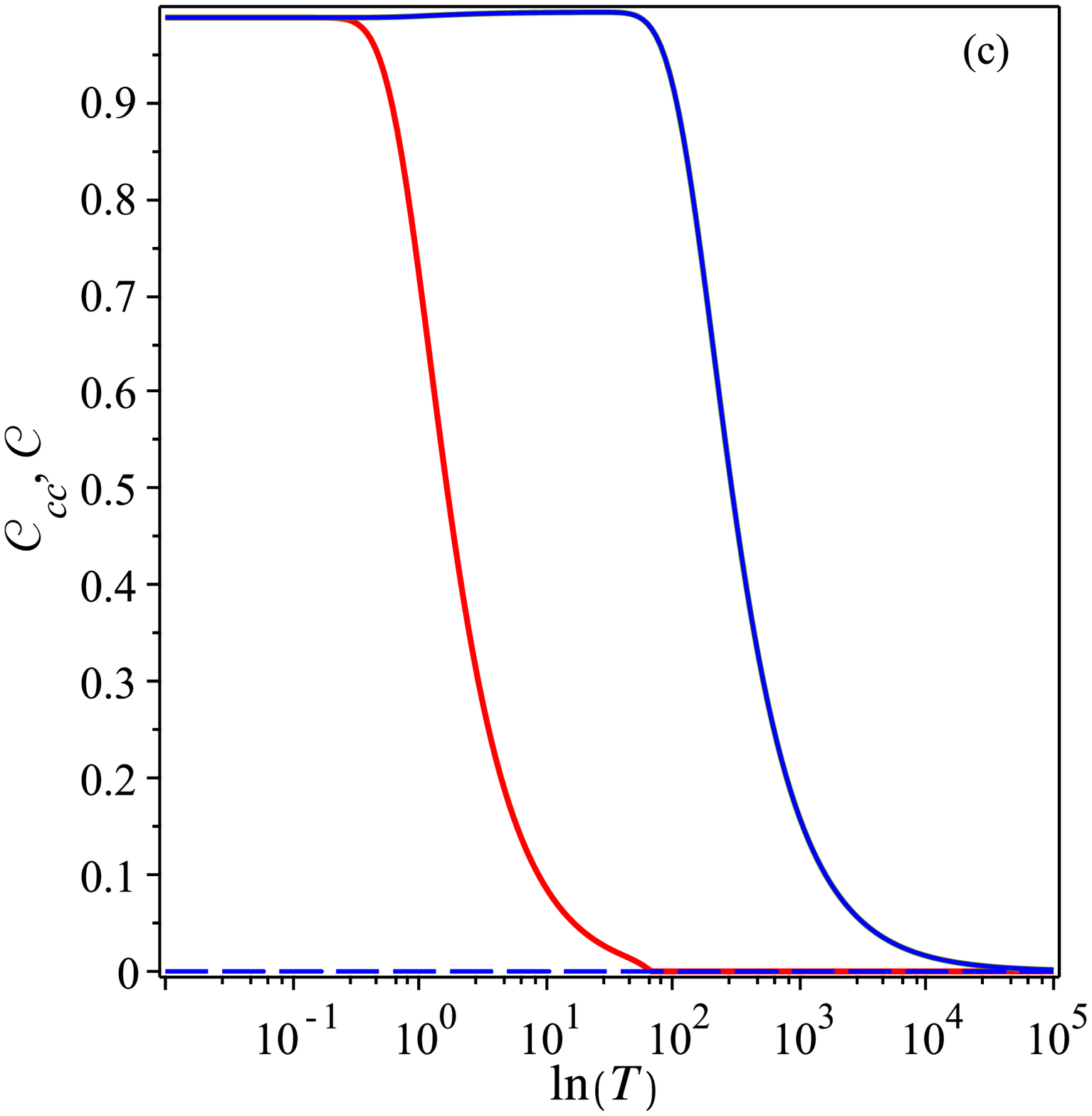}

\caption{\label{fig:Cc-C} Comparison between the correlated coherence $\mathcal{C}_{cc}$
(blue solid line) and the concurrence $\mathcal{C}$ (red solid line)
as a function of temperature $T$ in the logarithmic scale, for fixed
$\Delta_{1}=10$, $\Delta_{2}=15$, $V=16\Delta_{1}$ and $\varphi=0$.
(a) $\theta=0$, (b) $\theta=0.95\left(\frac{\pi}{4}\right)$, (c)
$\theta=\left(\frac{\pi}{4}\right)$.}
\end{figure}
\begin{figure}
\includegraphics[scale=0.4]{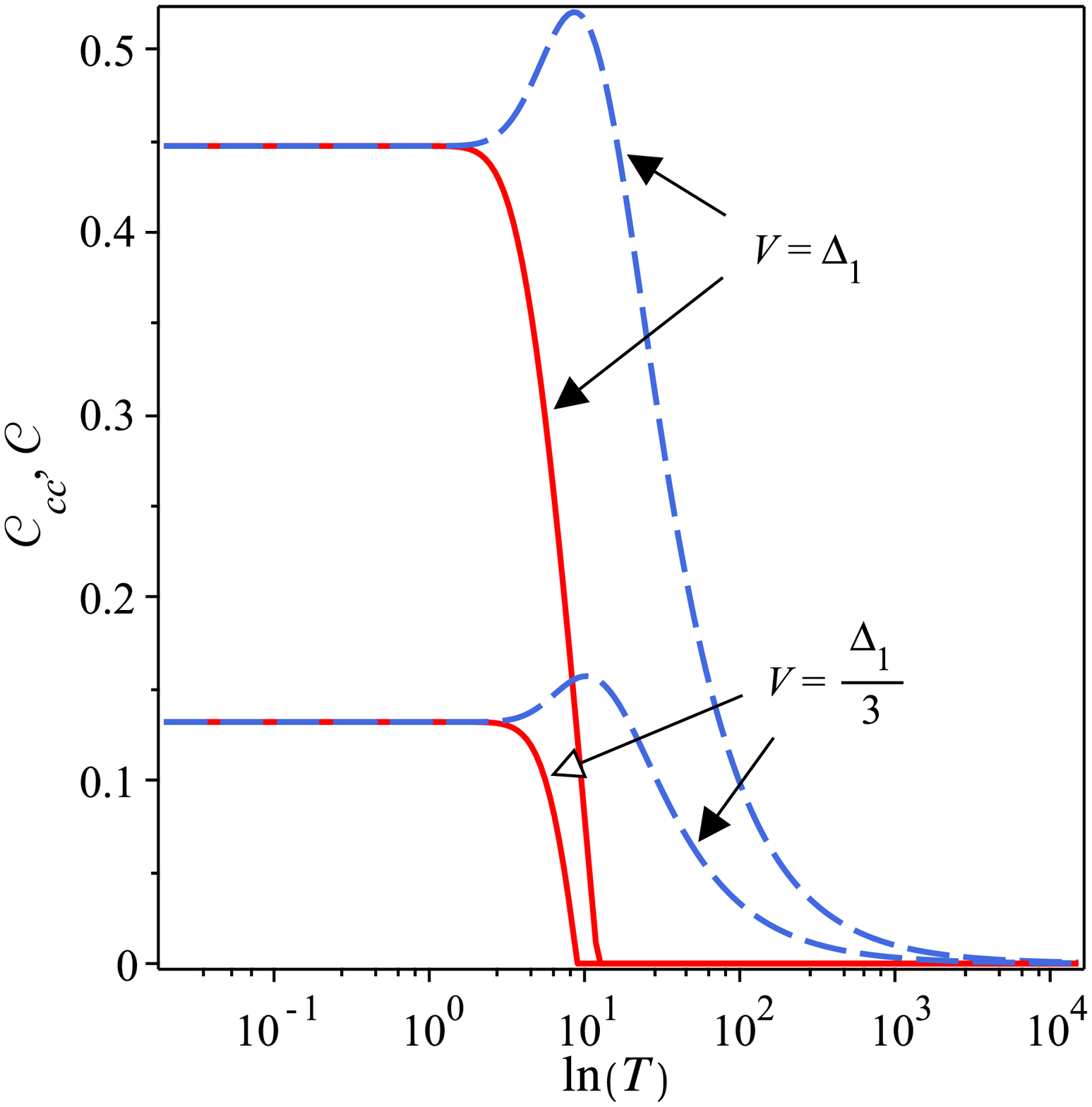}
\caption{\label{fig:6} Concurrence $\mathcal{C}$ (red solid curve) and the correlated coherence $\mathcal{C}_{cc}$ (blue dashed curve) versus $T$ in the
logarithmic scale for different values of Coulomb potential $V$ and
$\Delta_{1}=10$, $\Delta_{2}=15$.}
\end{figure}
In Fig. \ref{fig:Cc-C}, we compare the correlated coherence
and the concurrence versus the temperature for fixed tunneling coupling($\Delta_{1}=10$, $\Delta_{2}=15$), for a strong Coulomb potential $V=16\Delta_{1}$ and for different values of the parameter $\theta$. In these figures, we include the curves of total quantum coherence $\mathcal{C}_{l_{1}}(\rho_{AB})$
and the local quantum coherence $\mathcal{C}_{l_{1}}(\rho_{A})+\mathcal{C}_{l_{1}}(\rho_{B})$
for a better understanding the behavior of the quantum coherence and the quantum correlations. In Fig. \ref{fig:Cc-C}(a), the correlated coherence and the concurrence are depicted as a function of temperature $T$, on the basis of the eigenenergies which corresponds to the angle $\theta=0$ and to $\varphi=0$ in the transformation $U$(see Eq. \ref{eq:7}). These
curves show that, for $T\rightarrow0$, the correlated coherence $\mathcal{C}_{cc}$
(solid blue curve) is higher than the thermal entanglement $\mathcal{C}$(solid red curve). The difference between them are the untangled quantum correlations (quantum discord). One can see clearly that, as the temperature increases, the entanglement (red curve) decays up to threshold temperature $T\approx66.9$, while that, the total quantum coherence gradually decreases as the temperature increases. In Fig. \ref{fig:Cc-C}(b), we choose $\theta$ close to $\frac{\pi}{4}$, $\left(\theta=0.95\frac{\pi}{4}\right)$ and $\varphi=0$, so we can
see that, in this basis, the local quantum coherence (dashed blue curve)
is almost null. Then, it can be seen that the total quantum coherence (solid green curve) is, almost, entirely constituted by the correlated coherence for this particular choice of $\theta$. On the other hand, for high temperatures and after the concurrence and the local coherence has disappeared,
the total quantum coherence is composed solely by non-entangled quantum correlations. In Fig. \ref{fig:Cc-C}(c), the concurrence and the quantum coherence are analyzed for the incoherent basis $\theta=\frac{\pi}{4}$ and $\varphi=0$. It is interesting to see that there is an overlap between the total quantum coherence $\mathcal{C}_{l_{1}}(\rho_{AB})$ and the correlated coherence $\mathcal{C}_{cc}$. This is due to the chosen basis, which minimizes the correlated coherence. Moreover, by comparing
the results of concurrence $\mathcal{C}$ and the correlated coherence $\mathcal{C}_{cc}$, one can notice that the correlated coherence and the total of the thermal entanglement have the same order, in the low temperature region, which is consistent with the previous result \cite{chuan}. It is also observed that, as the temperature increases, the
entanglement decays and disappears for $T\approx66.9$, while that the correlated coherence increases slightly due to the thermal fluctuations, then decreasing monotonically as soon as the temperature $T$ increases.

Finally, in Fig. \ref{fig:6}, we plot both $\mathcal{C}_{cc}$ and $\mathcal{C}$ versus $T$, for fixed tunneling coupling $\Delta_{1}=10$, $\Delta_{2}=15$ and two different values of the Coulomb potential, namely, $V=\Delta_{1}$ and $V=\frac{\Delta_{1}}{3}$. In this figure, the correlated
coherence is analyzed in the incoherent basis for a local coherent $\theta=\frac{\pi}{4}$ and $\varphi=0$. As it can be clearly seen in the figure  \ref{fig:6}, the correlated coherence (blue dashed curve)
is equal to the concurrence (red curve) in the low
temperature region, i.e, the correlated coherence captures
the total thermal entanglement. As the temperature increases, the thermal fluctuations generate an increased quantum coherence, while the thermal entanglement decreases until vanishes at the threshold temperature, $T\approx12.24$ and $T\approx9.02$, for $V=\Delta_{1}$ and $V=\frac{\Delta_{1}}{3}$
respectively. Finally, the correlated coherence monotonously leads
to zero.
\section{Conclusions}
In summary, we have investigated the thermal entanglement and the correlated coherence in two coupled quantum dot molecules containing two excess electrons. Here, the correlated coherence was defined based on the recently formulated resource theory of coherence. In this model, we consider an isolated double quantum dot as a charge qubit and the proposed model was exactly solved. We obtained an explicit expression for the thermal density operator. This allows us to calculate the
thermal concurrence and the correlated coherence. We discussed in detail the effects of the tunneling coupling parameters, the Coulomb interaction between two electrons on DQDs as a function of temperature. Our results suggest that the Coulomb potential can be used to turn on the thermal entanglement and, then be tuned conveniently. This fact demonstrates the positive role of the Coulomb potential and enhancing both the thermal entanglement and the correlated coherence.

In addition, we found a direct connection between the entanglement and the quantum coherence. In particular, we reported that the correlated coherence measure is equal to the concurrence for low temperatures. Then, the thermal entanglement should be viewed as a particular form of a quantum coherence. Thus, here we consider $\theta=\frac{\pi}{4}$ and $\varphi=0$, which correspond to the incoherent basis for a local coherence. In
addition, the model showed a peculiar thermally-induce increase of
the correlated coherence due to the emergence of non-entangled quantum correlations, as the entanglement decrease. When $T$ is high enough, the quantum entanglement disappears as thermal fluctuation dominates the system.
\section{Acknowledgments}
This work was partially supported by CNPq, CAPES and Fapemig. M. R.
would like to thank CNPq grant 432878/2018-1.


\begin{thebibliography}{10}
\bibitem{Ben1} C. H. Bennett, G. Brassard, C. Crépeau, R. Jozsa,
A. Peres, W. K. Wootters, Phys. Rev. Lett. \textbf{70}, 1895 (1993).

\bibitem{Ben2} C. H. Bennett, H. J. Bernstein $\mathit{et\,al}$,
Phys. Rev. A \textbf{53}, 2046 (1996).

\bibitem{lamico} L. Amico, R. Fazio, A. Osterloh and V. Vedral, Rev.
Mod. Phys. \textbf{80}, 517 (2008).

\bibitem{peta} J. R. Peta, A. C. Johnson, J. M. Yaylor, E. A. Laird,
A. Yacoby, M. D. Lukin, C. M. Marcus, M. P. Hanson, A. C. Gossard,
Science, \textbf{309}, 2180 (2005).

\bibitem{press} D. Press, D. L. Thaddeus, Z. Bingyang, Y. Yamamoto,
Nature \textbf{456}, 218 (2008).

\bibitem{shin} a) G. Shinkai, T. Hayashi, T. Ota, T. Fujisawa, Phys.
Rev. Lett. \textbf{103}, 056802 (2009); b) T. Fujisawa, T. Hayashi, Y.
Hirayama, J. Vac. Sci. Technol. B \textbf{22}, 2035 (2004).

\bibitem{aus} D. G. Austing, T. Honda, K. Muraki, Y. Tokura, S. Tarucha,
Physica B \textbf{249}, 206 (1998).

\bibitem{so} Sophia E. Economou, Juan I. Climente, Antonio Badolato,
Allan S. Bracker, Daniel Gammon, Matthew F. Doty, Phys. Rev. B \textbf{86},
085319 (2012).

\bibitem{gor} J. Gorman, D. G. Hasko, D. A. Williams, Phys. Rev.
Letts. \textbf{95}, 090502 (2005).

\bibitem{benito} M. Benito, X. Mi, J. M. Taylor, J. R. Petta, Guido
Burkard, Phys. Rev. B, \textbf{9}6, 235434 (2017).

\bibitem{loss} D. Loss, D. P. DiVincenzo, Phys. Rev. A \textbf{57},
120 (1998).

\bibitem{an} B. D'Anjou, G. Burkard, Phys. Rev. B \textbf{100},
245427 (2019).

\bibitem{shi} Z. Shi, C. B. Simmons, J. R. Prance, John King Gamble,
Teck Seng Koh, Yun-Pil Shim, Xuedong Hu, D. E. Savage, M. G. Lagally,
M. A. Eriksson, Mark Friesen, S. N. Coppersmith, Phys. Rev. Lett.
\textbf{108}, 140503 (2012).

\bibitem{yang} Y. -C. Yang, S. N. Coppersmith, M. Friesen, Phys. Rev. A \textbf{101},
012338 (2020).

\bibitem{ita} T. Itakura, Y. Tokura, Phys. Rev. B \textbf{67}, 195320
(2003).

\bibitem{urda} M. Urdampilleta, A. Chatterjee, Ch. Ch. Lo, T. Kobayashi,
J. Mansir, S. Barraud, A. C. Betz, S. Rogge, M. F. Gonzalez-Zalba,
J. J. L. Morton, Phys. Rev. X \textbf{5}, 031024 (2015).

\bibitem{villas} J. M. Villas-Bôas, A. Govorov, Sergio E. Ulloa,
Phys. Rev. B \textbf{69}, 125342 (2004).

\bibitem{sanz} P. A. Oliveira, L. Sanz, Annl. Phys. \textbf{356},
244 (2015).

\bibitem{sza} B. Szafran, Phys. Rev. B \textbf{101}, 075306 (2020).

\bibitem{fan} F. F. Fanchini, L. K. Castelano, A. O. Caldeira, New.
J. Phys. \textbf{12}, 073009 (2010).

\bibitem{qin} Xiao-Ke Qin, Europhys. Lett. \textbf{114}, 37006 (2016).

\bibitem{borge} H. S. Borges, L. Sanz, J. M. Villas-Bôas, O. O. Diniz
Neto, A. M. Alcalde, Phys. Rev. A \textbf{85}, 115425 (2012).

\bibitem{sou} F. M. Souza, P. A. Oliveira, L. Sanz, Phys. Rev. A \textbf{100}, 042309 (2019).

\bibitem{choo} a) K. W. Choo, L. C. Kwek, Phys. Rev. B, \textbf{75},
205321 (2007); b) F. D. Pasquale, G. Georgi, S. Pagonelli, Phys. Rev.
Lett, \textbf{93}, 120502 (2004).

\bibitem{gia} A. Purkayastha, G. Guarnieri, M. T. Mitchison, R. Filip,
J. Goold, npj Quantum Inf., \textbf{6}, 27 (2020).

\bibitem{rao} D. D. B. Rao, S. Gosh, P. K. Panigrahi, Phys. Rev.
A \textbf{78}, 042328 (2008).

\bibitem{chot} L. Chotorlishvili, A. Gudyma, J. Watzel, A. Ernst, J. Berakdar, Phys. Rev.
B \textbf{100}, 174413 (2019).

\bibitem{divi} D. P. DiVincenzo, D. Bacon, J. Kempe, G. Burkard and
K. B. Whaley, Nature, \textbf{408}, 339 (2000).

\bibitem{bose} S. Bose, Phys. Rev. Lett. \textbf{91}, 207901 (2003).

\bibitem{stre} A. Streltsov, E. Chitambar, S. Rana, M. N. Bera, A.
Winter, M. Lewenstein, Phys. Rev. Lett. \textbf{116}, 240405 (2016).

\bibitem{stre1} A. Streltsov, G. Adesso, M. B. Plenio, Rev. Mod.
Phys. \textbf{89}, 041003 (2017).

\bibitem{fro} F. Fr$\ddot{\textrm{o}}$wis, W. D$\ddot{\textrm{u}}$r,
Phys. Rev. Lett. \textbf{106}, 110402 (2011).

\bibitem{gio} V. Giovannetti, S. Lloyd, L. Maccone, Science \textbf{306},
1330 (2004).

\bibitem{brandao} a) F. G. S. L. Brandão, M. Horodecki, J. Oppennheim,
J. M. Renes, R. W. Spekkens, Phys. Rev. Lett. \textbf{111}, 250404
(2013); b) M. Lostaglio, D. Jennings, T. Rudolph, Nat. Commun. \textbf{6},
6383 (2015). 

\bibitem{lan} J. P. Santos, L. C. Céleri, G. T. Landi, M. Paternostro,
Nat. Quant. Inf. \textbf{5}, 23 (2019).

\bibitem{baum} T. Baumgratz, M. Cramer, M. B. Plenio, Phys. Rev.
Lett. \textbf{113}, 140401 (2014).

\bibitem{Hu} M. L. Hu, X. Hu, J. C. Wang, Y. Peng, Y. R. Zhang, H.
Fan, Phys. Rep. \textbf{762}, 1 (2018).

\bibitem{tan} K. C. Tan, H. Kwon, C-Y. Park, H. Jeong, Phys. Rev.
A \textbf{94}, 022329 (2016).

\bibitem{tri} T. Kraft, M. Piani, J. Phys. A: Math. Theor. \textbf{51},
414013 (2018).

\bibitem{ollivier} a) H. Ollivier, W. H. Zurek, Phys. Rev. Lett. \textbf{88},
017901 (2012); b) D. Girolami, G. Adesso, Phys. Rev. A,\textbf{83}, 052108 (2011).

\bibitem{wer} T. Werlang, G. Rigolin, Phys. Rev. A \textbf{81},
044101 (2010).

\bibitem{yue} X. -L. Wang, Q. -L. Yue, Ch. -H. Yu, F. Gao, S. -J. Qin, Sci. Rep. \textbf{7}, 12122 (2017).

\bibitem{ma} J. Ma, B. Yadin, D. Girolami, V. Vedral, M. Gu, Phys. Rev. Letts. \textbf{116},
160407 (2016).

\bibitem{zhang} a) G. F. Zhang, S. S. Li, Phys. Rev. A, \textbf{72}, 034302
(2005); b) M. C. Amesen, S. Bose, V. Vedral, Phys. Rev. Lett. \textbf{87},
017901 (2001).

\bibitem{cheng} W. W. Cheng, X. Y. Wang, Y. B. Sheng, L. Y. Gong,
S. M. Yhao, J. M. Liu, Sci. Rep. \textbf{7}, 42360 (2017).

\bibitem{wootters} W. K. Wootters, Phys. Rev. Lett. \textbf{80},
2245 (1998).

\bibitem{chuan} K. Ch. Tan, H. Jeong, Phys. Rev. Lett. \textbf{121},
220401 (2018).
\end{thebibliography}
\end{document}